\documentclass[showpacs,preprintnumbers,amsmath,amssymb,twocolumn,superscriptaddress,prb]{revtex4}
\usepackage{times}
\usepackage{graphicx}
\usepackage{amsfonts}
\usepackage{amsmath, amsthm, amssymb}

\newcommand{\ve}[1]{\mathbf{#1}}

\DeclareMathOperator{\diag}{diag} 

\newcommand{\vk}{\ve{k}} 
\newcommand{\vp}{\ve{p}} 

\newcommand{\e}[1]{\mathrm{e}^{#1}}

\def\i{\mathrm{i}} 

\newcommand{\ie}{\textit{i.e. }}
\newcommand{\eg}{\textit{e.g. }}
\newcommand{\etal}{\emph{et al.}}

\begin{document}
\title[Josephson effect in thin-film superconductor/insulator/superconductor junctions with misaligned 
in-plane magnetic fields]{Josephson effect in thin-film superconductor/insulator/superconductor 
junctions with misaligned in-plane magnetic fields}
\author{J. Linder}
\affiliation{Department of Physics, Norwegian University of
Science and Technology, N-7491 Trondheim, Norway}
\author{A. Sudb{\o}}
\affiliation{Department of Physics, Norwegian University of
Science and Technology, N-7491 Trondheim, Norway}
\affiliation{Center for Advanced Study, The Norwegian Academy of 
Science and Letters, N-0271 Oslo, Norway}
\date{Received \today}
\begin{abstract}
We study a tunnel junction consisting of two thin-film $s$-wave superconductors separated by a thin, 
insulating barrier in the presence of misaligned in-plane exchange fields. We find an interesting 
interplay between the superconducting phase difference and the relative orientation of the exchange 
fields, manifested in the Josephson current across the junction. Specifically, this may be written 
$I_\text{J}^\text{C} = (I_0+I_m ~ \cos\varphi)~ \sin\Delta\theta$, where $I_0$ and $I_m$ are constants, 
and $\varphi$ is the relative orientation of the exchange fields while $\Delta\theta$ is the 
superconducting phase difference. Similar results have recently been obtained in other S/I/S 
junctions coexisting with helimagnetic or ferromagnetic order. We calculate the superconducting 
order parameter self-consistently, and investigate quantitatively the effect which the misaligned 
exchange fields constitute on the Josephson current, to see if $I_m$ may have an appreciable effect 
on the Josephson current. It is found that $I_0$ and $I_m$ become comparable in magnitude at sufficiently 
low temperatures and fields close to the critical value, in agreement with previous work. From our 
analytical results, it then follows that the Josephson current in the present system may be controlled 
in a well-defined manner by a rotation of the exchange fields on both sides of the junction. We 
discuss a possible experimental realization of this proposition.
  \end{abstract}

\maketitle

\section{Introduction}
The study of physical effects that arise due to an interplay between superconductivity (SC) and ferromagnetism (FM) has 
grown considerably over the last decade (see Refs.~\onlinecite{bergeret, buzdin} and references therein). Much effort has been devoted to obtaining a better understanding of the exotic phenomena that may appear in heterostructures of superconductors/ferromagnets. To mention a few of these, it is natural to highlight the study of $\pi$-junctions, both theoretically \cite{pijunctionst} and experimentally \cite{pijunctionse}, and the proximity effects giving rise to induced SC correlations in normal metals/half-metals/FM metals \cite{eschrig2003, bergeret2001} as prime examples of the potential that lies within this field of research. Also, quite recently, the coexistence of SC and FM in the same material was discovered in \cite{aoki, saxena} UGe$_2$ and URhGe, and possibly \cite{pfleiderer1, pfleiderer2} also in ZrZn$_2$. Such ferromagnetic superconductors (FMSC) display simultaneously multiple broken symmetries 
[SU(2) and U(1)], an interesting property that may be exploited in terms of dissipationless quantum transport of spin and/or charge between such materials \cite{kulic,eremin,gronsleth}. 
\par
Besides the interest from a fundamental physics point of view, transport properties in SC/FM heterostructures currently attract much attention, since it is hoped that the new physics that emerges in this type of systems may be useful for applications in nanotechnology and spintronics \cite{zutic}. The discoveries of unconventional superconductors displaying $d$-wave singlet \cite{harlingen}, $p$-wave triplet \cite{maenoRMP}, and even mixed singlet-triplet SC pairing symmetries \cite{gorkov, bauer}, offers the theoretician a true goldmine in terms of rich physics and opportunities to explore. In the present paper, however, we will be concerned with a system of two thin-film spin-singlet $s$-wave superconductors separated by a thin, insulating barrier in the presence of misaligned in-plane exchange field. This would be equivalent to a F/S/I/S/F system assuming that the S/F bilayer is thin and thus may be represented by a BCS superconductor in the presence of a homogenous magnetic field \cite{li2002}. Indeed, for superconducting films of thickness $t < \xi \ll \lambda$, where $\xi$ is the coherence length (average size of the Cooper pairs) and $\lambda$ is the magnetic field penetration depth, a magnetic field which is applied in the plane of the film will penetrate it practially uniformly. In this case, the Meissner effect-response of the superconductor is incomplete, such that the screening currents are minimal \cite{meservey}. Since orbital effects are suppressed in such a geometry, the critical field is determined by the paramagnetic limitation. Such type of systems have been considered earlier \cite{ambegaokar, fiske, likharev, bergeretJOS}. Nevertheless, 
we hope to shed some light on a matter which has not been investigated extensively in such systems: manipulating a supercurrent of spin and/or charge by controlling a misalignment of magnetic fields present on both sides of the barrier. Such a proposition was first made by Kulic and Kulic \cite{kulic} in 2001 (albeit in a physically completely different system), who derived an expression for the Josephson current over a junction separating two spin-singlet superconductors with spiral magnetic order. It was found that the supercurrent could be controlled by adjusting the relative orientation of the exchange field on both sides of the junction, 
a finding that quite remarkably suggested a way of tuning a supercurrent in a well-defined manner from \eg a 0- to $\pi$-junction. However, from an experimental point of view such states are
very hard to realize. Moreover it is extremely difficult, if not impossible, to control the magnetization
misalignment across the tunneling junction. 
Later investigations made by Eremin, Nogueira, and Tarento \cite{eremin} considered a similar system as Kulic and Kulic, namely two Fulde-Ferrel-Larkin-Ovchinnikov (FFLO) superconductors \cite{fflo} coexisting with helimagnetic order \cite{varma}. Recently, the same effect was found to exist in a FMSC/I/FMSC junction as 
shown by Gr{\o}nsleth \etal \cite{gronsleth}, a system which presumably has a much better
potential for being realized. 
\par
In the present paper, we show that a similar effect may be realized by applying misaligned in-plane exchange 
fields in a thin-film F/S/I/S/F junction, where S represents an s-wave thin film superconductor in
an external magnetic field provided by F (a ferromagnet).
 Such a system should be possible to realize experimentally.
We derive the linear-response expression for the Josephson current 
within the Matsubara formalism, and solve for the SC order parameter self-consistently, thereafter providing 
numerical results for the supercurrent that arises in the system for arbitrary misalignment of the magnetic 
field across the junction. We investigate under what experimental
conditions the predicted modulation of the total Josepshon current is most easily
observed. We also suggest an experimental setup to test these predictions.
\par
\indent This paper is organized as follows. In Sec. \ref{sec:model}, we establish our model and the 
formulation to be used throughout the paper, and solve for the SC order parameter self-consistently. 
The Josephson current is calculated within the tunneling Hamiltonian formalism in Sec. \ref{sec:jos}. 
Our main findings for the numerical values of the parameters that determine the modulation of the 
Josephson current as a function of the twist in the orientation of the exchange fields on both sides 
of the junction, are presented in Sec. \ref{sec:res} with a discussion given in Sec. \ref{sec:dis}. 
In this section, we also provide a description 
of a possible heterostructure for realizing the physical situation we describe in this paper. Specifically,
we suggest how one may be able to physically misalign an external field across the tunneling junction (by 
an arbitrary amount). Finally, we summarize our results in Sec. \ref{sec:sum}, and reemphasize what 
our new findings are compared to previous results.

\section{Model and formulation}\label{sec:model}
The total Hamiltonian $H$ for a system consisting of two superconductors separated by an 
insulating layer in the presence of an in-plane exchange field can be written as \cite{cohen1} 
$H = H_\text{L} + H_\text{R} + H_\text{T}$, where L and R represent the individual 
superconductors on each side of the tunneling junction, and $H_\text{T}$ describes 
tunneling of particles through the insulating layer separating the two superconductors. 
At the level of mean-field theory the individual superconductors are described by
\begin{equation}
  \label{eq:1}
  H = H_0 + \sum_{\vk}\varphi_{\vk}^\dag
  \cal{A}_{\vk}\varphi_{\vk},  
\end{equation}
where $H_0$ is given by
\begin{align}\label{eq:A}
H_0 &= \sum_{\vk}\xi_{\vk} -
\sum_{\vk} \Delta^\dag
b_{\vk} + \frac{|\mathbf{H}|^2}{2\mu_0},\notag\\
{\cal{A}}_\vk &= 
\begin{pmatrix}
\xi_{\vk} - h & \Delta\e{\i\theta} \\
 \Delta\e{-\i\theta} &- \xi_{\vk} -h  \\
\end{pmatrix},
\end{align}
Here, $\vk$ is the electron momentum, $\xi_{\vk} = \varepsilon_{\vk} - \mu$,
$\sigma=\uparrow,\downarrow=\pm 1$, $\mu$ is the chemical potential (which at $T=0$ 
is completely equivalent to the Fermi energy), $\mathbf{H}$ is the magnetic field, 
$h$ is the exchange energy, $\mu_0$ is the magnetic permeability,  while 
$\Delta\e{\i\theta}$ is the superconducting order parameter and
$b_{\vk} = \langle c_{-\vk\downarrow}c_{\vk\uparrow}\rangle$
denotes the two-particle operator expectation value.
Eq. (\ref{eq:A}) is valid for an $s$-wave superconductor with 
an in-plane exchange field giving rise to an exchange interaction. 
\indent At this point, some comments are in order. We assume that no vortices are present in 
the system. This puts  limitations on the dimension of the thin-film. Our assumption 
of a homogenous exchange field in the superconductors can only be justified given 
that the thickness of the film is smaller than \cite{meservey} both the penetration 
depth $\lambda$ and coherence length $\xi$. The physical reason for this is that 
an externally applied in-plane magnetic field is found to penetrate the superconductor 
without creating vortices as long as there is no room for the 
vortices, which typically have a diameter of ${\cal{O}}(\xi)$. 
This amounts to a thickness of order $10$ nm, which is well 
within reach of current experimental techniques. \\
\indent Moreover, we will neglect phase-fluctuations and amplitude fluctuations in 
the superconducting order parameter in this paper. Amplitude-fluctuations may safely be neglected \cite{tesanovic1999,nguyen1999}. In a strong type-II
superconductor, neglecting critical fluctuations (which are
transverse phase-fluctuations, or equivalently vortices) is certainly not valid close enough to the
normal metal - superconductor transition \cite{tesanovic1999,nguyen1999}. 
In type-II superconductors, neglect of critical fluctuations
is expected to be reasonable provided we 
are outside the critical region, which is expected to be quite narrow around the 
critical temperature and critical field unless the superconductors are
of the extreme type-II \cite{tesanovic1999,nguyen1999}.  
 In deep type-I superconductors, the mean field 
approximation is expected to be excellent in any case, since the phase transition 
in such systems is of first order \cite{hlm1974,bartholomew1983,mo2002}.

\begin{figure}[h!]
\centering
\resizebox{0.48\textwidth}{!}{
\includegraphics{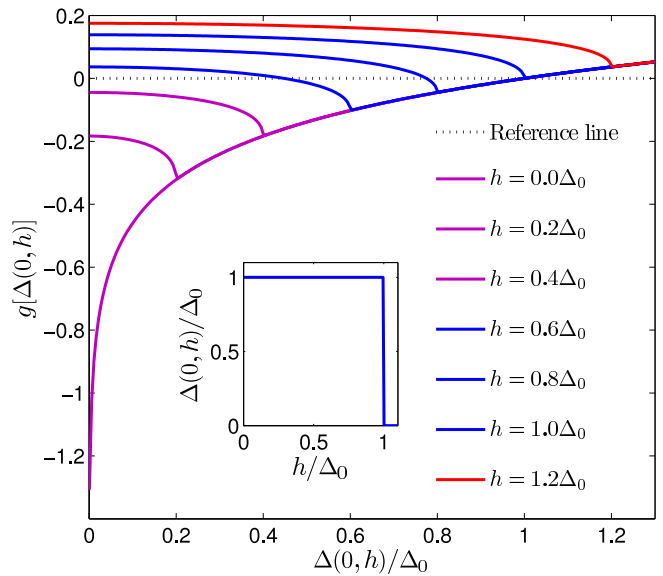}}
\caption{(Color) Plot of the function $g[(\Delta(0,h)]$ given by Eq. (\ref{eq:selfconsist}) to 
illustrate the possible solutions for the gap, given by where the curves intersect the dotted 
line. When $h/\Delta_0>0.5$, there is more than one solution to the gap equation, but only one 
of these are stable. As shown in the inset, where we have plotted the field dependence of this 
stable solution, a first order phase transition to the normal state is present at zero temperature.  }
\label{fig:1}
\end{figure}

\begin{figure}[h!]
\centering
\resizebox{0.48\textwidth}{!}{
\includegraphics{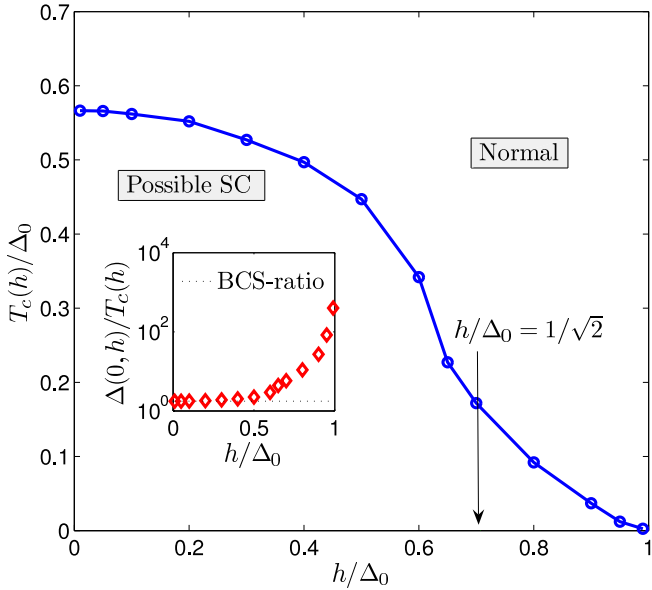}}
\caption{(Color) The phase diagram in the $h$-$T$ plane for a superconductor in the presence of an exchange field. A non-zero solution for the gap exists under the dotted line, indicating a possible SC phase. The exact regime where SC is energetically favored over the normal state was studied in Ref.~\onlinecite{li2002}, see their Fig. 1. Since the phase transition is first order, note that the ratio $\Delta(T,h)/T_c(h)$ is not constant as in the pure BCS case, as shown in the inset. }
\label{fig:2}
\end{figure}

In Eq. (\ref{eq:1}), our basis is 
\begin{equation}
\varphi_{\vk} =
(c_{\vk\uparrow}~c_{\mathbf{-\vk}\downarrow}^\dag)^{\text{T}},
\end{equation}
 where $\{c_{\vk\sigma},
c_{\vk\sigma}^\dag\}$ are annihilation and creation fermion operators with momentum $\vk$ and spin $\sigma$. By diagonalizing Eq. (\ref{eq:A}) through ${\cal{A}}_{\vk} = P_{\vk}D_{\vk}P_{\vk}^\dag$, Eq. (\ref{eq:1}) turns into 
\begin{equation}\label{eq:H2}
H = H_0 +
\sum_{\vk}\widetilde{\varphi}^\dag_{\vk}D_{\vk}\widetilde{\varphi}_{\vk},
\end{equation}
where the diagonal matrix reads $D_{\vk} = \diag(E_{\vk\uparrow}, E_{\vk\downarrow})$, and the basis $\widetilde{\varphi}_\vk$ consists of new fermion operators according to 
\begin{equation}
\widetilde{\varphi}_{\vk} = P_{\vk}^\dag \varphi_{\vk}
= ({\cal{C}}_{\vk\uparrow} ~{\cal{C}}_{-\vk\downarrow}^\dag)^{\text{T}}.
\end{equation}
Upon defining the auxiliary quantity
\begin{equation}
R_\vk = \frac{\Delta}{\xi_\vk + \sqrt{\xi_\vk^2 + \Delta^2}},
\end{equation}
the diagonalization matrix may be written as 
\begin{align}
P_\vk &= N_\vk\begin{pmatrix}
1 & -R_\vk\e{\i\theta}\\
R_\vk\e{-\i\theta} & 1 \\
\end{pmatrix},\notag\\
N_\vk &= 1/\sqrt{1 + R_\vk^2}.
\end{align}
We find that the energy eigenvalues may be written as
\begin{align}
E_{\vk\sigma} = \sigma\sqrt{\xi_\vk^2 + \Delta^2} - h.
\end{align}
Concerning ourselves with $s$-wave pairing ($\vk$-independent gap), we note that $E_{\vk\sigma}=E_{-\vk\sigma}$, which 
allows us to recast Eq. (\ref{eq:H2}) into the form
\begin{equation}
H = H_0 - \sum_\vk E_{\vk\downarrow} + \sum_{\vk\sigma} \sigma E_{\vk\sigma} {\cal{C}}_{\vk\sigma}^\dag {\cal{C}}_{\vk\sigma}.
\end{equation}
The self-consistent gap equations are derived from the free energy given by
\begin{equation}\label{eq:F}
F = H_0 - \sum_{\vk} E_{\vk\downarrow} - \frac{1}{\beta}\sum_{\vk\sigma}\text{ln}(1 + \e{-\beta\sigma E_{\vk\sigma}}).
\end{equation}
yielding the self-consistency equation
\begin{align}
\label{eq:selfconsist}
g(\Delta) \equiv 1 - \frac{c}{2} \int^{\omega_0}_{-\omega_0} \text{d}\xi \Big\{ \frac{1-f[E_\uparrow(\xi)]-f[-E_\downarrow(\xi)]}{\sqrt{\xi^2+\Delta^2}}\Big\} = 0,
\end{align}
where the weak-coupling constant $c=VN(0)$ is set to 0.2 hereafter, while $\omega_0$ is arbitrarily 
set to 1\% of the Fermi energy, \ie $\mu/100$, which corresponds to $\omega_0/\Delta \approx 70$, 
which essentially is equivalent to $\omega_0/\Delta \to \infty$. ($\omega_0/\Delta \approx 10$ suffices 
to achieve this limit in the quantities we consider in this paper). 
In the limit of zero exchange field, $h\to 0$, the well-known result (see \eg Ref.~\onlinecite{likharev}) 
is obtained. The Fermi-Dirac distribution functions entering in Eq. (\ref{eq:selfconsist}) are given as 
$f(\xi) = 1/(1+\e{\beta \xi})$ where $\beta$ is inverse temperature. We have introduced the usual 
simplification of a pairing potential that is attractive in a small energy interval around Fermi-level
\begin{equation}
V_{\vk\vk'\alpha\beta} = -V \text{ for } |\xi_{\vk(\vk')} - \mu|<\omega_0,
\end{equation}
with $(V>0)$, and zero otherwise. Here, $\omega_0$ is a typical frequency cutoff defining the spectral 
width of the bosons responsible for the pairing. We do not further specify what these bosons are. 
Eq. (\ref{eq:selfconsist}) will be the governing equation for the gap $\Delta = \Delta(T,h)$ at an 
arbitrary temperature and arbitrary in-plane exchange field. The orbital effect from the exchange 
field in this configuration is suppressed, since the electrons are restricted from moving in the $\hat{\mathbf{z}}$-direction due to the thin-film structure. \\
\indent The order parameter may now be solved for numerically, by integrating the gap equation Eq. (\ref{eq:selfconsist}). Consider first the zero temperature case, where we have plotted the dependence 
of $g (\Delta)$ on $h$ in Fig. \ref{fig:1}, such that the possible solutions are identified by locating 
the intersection with the dotted line defined by $g(\Delta)=0$. In agreement with previous results \cite{fflo}, we find that for $h/\Delta_0<0.5$ there is a unique solution of $\Delta(0,h)$ that satisfies $g[\Delta(0,h)] = 0$, 
while another solution $\Delta(0,h)<\Delta_0$ is present for $0.5<h/\Delta_0<1.0$. However, this has 
been found to be unstable, such that we will only consider the solution for the largest gap \cite{fflo}. 
In this case, one may simply write 
\begin{align}
\Delta(0,h) = \left\{ \begin{array}{ll}
\Delta_0 &\text{if } h<\Delta_0 \\
0 &\text{if } h>\Delta_0. \\
\end{array} \right.
\end{align}
In the inset of Fig. \ref{fig:1}, we have plotted the field-dependence of the stable solution 
$\Delta(0,h)$. As shown, there is a first order phase transition at $h=\Delta_0$ whereas the 
gap remains independent on $h$ for $h<\Delta_0$.
Consider now the dependence of the critical temperature as a function of $h$, illustrated in Fig. \ref{fig:2}. Effectively, the $T_c$ vs. $h$ curve gives the phase diagram of a superconductor with an in-plane exchange field. 
Note that although a non-zero solution for $\Delta$ exists under the dotted line in Fig. \ref{fig:2}, one must 
turn to free energy considerations in order to determine whether the normal state or superconducting state is 
favored. Such a study was undertaken in Ref.~\onlinecite{li2002} (see their Fig. 1). The 
Clogston-Chandrasekhar critical field $h = \Delta_0/\sqrt{2}$ at $T=0$ is also given in the Fig. 
\ref{fig:2} \cite{clogston, chandrasekhar}. In the present paper, we will be concerned with the field 
dependence of the physical quantities, and thus choose five representative temperatures 
(see Tab. \ref{tab:list}) at which the SC state is indeed the thermodynamical state favored, 
as given by Ref.~\onlinecite{li2002}.

\begin{table}[h!]
\centering{
\caption{Critical field at the five representative temperatures we will study \cite{li2002}.}
	\label{tab:list}
	\vspace{0.15in}
	\begin{tabular}{cc}
			  	  	 \hline
		  Temperature $T/\Delta_0$ \hspace{0.1in}	& Critical field $h_c/\Delta_0$  \\
		  	  	 \hline
	  	 \hline
		  0.001 \hspace{0.1in}	& 0.70  \\
		 0.1 \hspace{0.1in}	& 0.68\\
		  0.2 \hspace{0.1in}	& 0.65  \\
		  0.3 \hspace{0.1in}	& 0.52  \\
		 0.4 \hspace{0.1in}	& 0.35 \\
	  	 \hline
	  	 \hline
	\end{tabular}}
\end{table}

\begin{figure}[h!]
\centering
\resizebox{0.48\textwidth}{!}{
\includegraphics{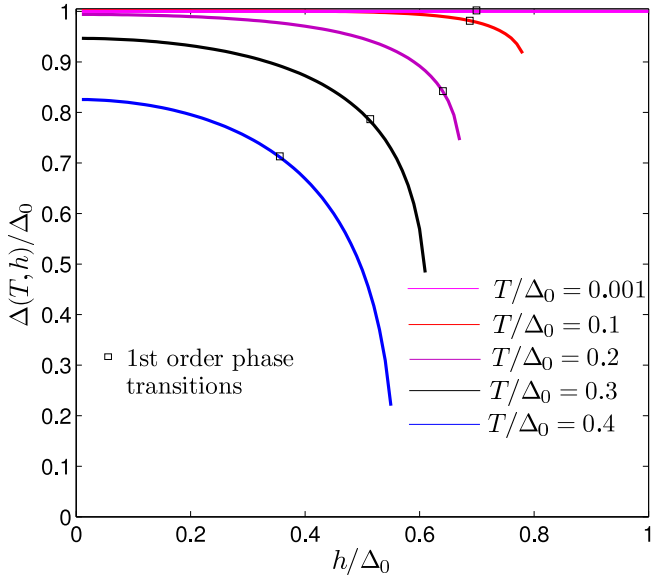}}
\caption{(Color) Field dependence $h$ of the superconducting order parameter $\Delta = \Delta(T,h)$ 
at finite temperatures. The sudden end of the curves clearly indicates a sharp drop in the gap, 
indicating a discontinuous nature of the normal metal-superconductor phase transition. }
\label{fig:3}
\end{figure}

Finally, we give a plot of the field dependence of $\Delta$ at finite temperatures, illustrated in Fig. \ref{fig:3}. It is seen that the phase transition at the critical field remains discontinuous also at finite temperatures \cite{sarma, maki1, maki2}.

\section{Josephson current}\label{sec:jos}
In order to calculate the Josephson charge-current over the junction, we make use of the equilibrium 
Matsubara Greens-function formalism at finite temperatures (see \eg Ref.~\onlinecite{mahan}). Since we 
are interested in misaligned exchange fields on both sides of the junction, we will use different 
quantization axes on the left and right side of the barrier. By including the Wigner $d$-function 
\cite{wigner}, one may then account for the fact that an $\uparrow$-spin on one side of the junction 
is not the same as an $\uparrow$-spin on the other side. Defining
\begin{equation}
\mathcal{D}(\varphi) = \begin{pmatrix}
\cos(\varphi/2) & -\sin(\varphi/2)\\
\sin(\varphi/2) & \cos(\varphi/2)\\
\end{pmatrix},
\end{equation}
the tunneling Hamiltonian of the present system may be written as
\begin{equation}
H_\text{T} = \sum_{\vk\vp\sigma\sigma'} [\mathcal{D}(\varphi)]_{\sigma\sigma'} \mathcal{T}_{\vk\vp} c_{\vk\sigma}^\dag d_{\vp\sigma'} + \text{h.c.}
\end{equation}
Above, $c_{\vk\sigma}$ designate fermion operators on the right side of the junction, while $d_{\vp\sigma}$ 
represents fermion operators on the left side of the junction, and $\mathcal{T}_{\vk\vp}$ is the tunneling 
probability amplitude. The Josephson charge-current is now defined as
\begin{equation}
I_\text{J}(t) = -e\Big\langle \frac{\text{d}N_\text{L}(t)}{\text{d}t}\Big\rangle,
\end{equation}
where the time derivative of the number operator is given by
\begin{equation}
\frac{\text{d}N_\text{L}(t)}{\text{d}t} = \i\e{\i H't}[H_\text{t}, N_\text{L}] \e{-\i H't}.
\end{equation}
We have defined $H' = H_\text{L}+H_\text{R}$, and only taken into account the contribution from the tunneling Hamiltonian to the time-derivative. In this way, the calculated current will only consist of processes corresponding 
to physical transport across the junction and not any additional contributions originating from a lack of particle conservation number on each side of the junction, respectively. The procedure to obtain $I(t)$ is now fairly straight-forward, and may be reviewed in \eg  Refs.~\onlinecite{kulic,eremin,gronsleth,linderprb2007}. We find 
that at zero applied voltage, the Josephson-current is time-independent and reads
\begin{equation}\label{eq:main}
I_\text{J} = (I_0 + I_m\cos\varphi)\sin\Delta\theta,
\end{equation}
where $\varphi$ is the relative orientation of the exchange fields and $\Delta\theta$ is the superconductivity phase difference across the junction. This establishes a Josephson current which may be controlled through an adiabatic rotation of misaligned exchange fields in a planar S/I/S system, or equivalently an F/S/I/S/F layer. While it is not clear how the exchange field could be experimentally controlled in a well-defined manner in junctions with BCS \cite{kulic} and FFLO \cite{eremin} superconductors coexisting with helimagnetic order, where this effect has been discussed previously \cite{kulic, eremin}, we will proceed to show that experimental verification of this type of effect should be more feasible in the present system. The amplitudes entering in Eq. (\ref{eq:main}) read
\begin{align}
I_0 = 2e\mathcal{T}^2 \sum_{\vk\vp} N_\vk^2R_\vk N_\vp^2R_\vp F_{\vk\vp}^+,\notag\\
I_m = 2e\mathcal{T}^2 \sum_{\vk\vp} N_\vk^2R_\vk N_\vp^2R_\vp F_{\vk\vp}^-,
\end{align}
where $\mathcal{T} = |\mathcal{T}_{\vk\vp}|$ is the tunneling amplitude (see discussion below) and
\begin{align}
F_{\vk\vp}^\pm = \sum_{\alpha\beta} \alpha\beta\Big[& \frac{f(E_{\vk\alpha})-f(E_{\vp\beta})}{E_{\vk\alpha}-E_{\vp\beta}}\notag\\
&\pm\frac{1 - f(E_{\vk\alpha})-f(E_{\vp\beta})}{E_{\vk\alpha} +E_{\vp\beta}}\Big].
\end{align}
Note that when the exchange field vanishes, we have that $F_{\vk\vp}^{-}=0$, such that
$I_m=0$. In general, therefore, for weak exchange fields we expect that 
$I_m \ll I_0$. Hence, an appreciable amount of modulation of the
total Josephson current $I_J$ by a twist in the magnetization across the
junction will require a certain amount of fine tuning. We will detail this below. 
\section{Results}\label{sec:res}
We now consider in more detail the Josephson current as a function of both temperature and twist in the exchange fields upon insertion of the self-consistent solutions of $|\Delta(T,h)|$ into the expression for the Josephson current, Eq. (\ref{eq:main}). To this end, we replace summation over momenta by integration over energies by means of the formula
\begin{equation}
\frac{1}{N}\sum_\vk \mathcal{F}_\vk = \int\int \text{d}\varepsilon \text{d}\Omega N(\varepsilon,\Omega) \mathcal{F}(\varepsilon,\Omega),
\end{equation}
where $\int \text{d}\Omega$ corresponds to an angular integration over a constant sheet of energy $\varepsilon$ in momentum space, $N(\varepsilon,\Omega)$ is the angularly resolved density of states, and $\mathcal{F}(\varepsilon,\Omega) = \mathcal{F}[\vk(\varepsilon,\Omega)]$ is an arbitrary function. In general, it is necessary to specifiy the nature of the tunneling matrix element in some detail, since the crude approximation $|\mathcal{T}_{\vk\vp}|^2 = \mathcal{T}^2$ may lead to unphysical results \cite{bruder}. A plausible conjecture for the tunneling matrix element should incorporate two key elements: \textit{i)} quasiparticles moving perpendicularly towards the junction should have a higher probability of tunneling than quasiparticles moving parallell to it, and \textit{ii)} the direction of momentum should be conserved in the tunneling-process, \ie a right-moving quasiparticle on the left side of the junction should only tunnel into a right-moving quasiparticle on the right side of the junction, and vice versa. However, due to isotropic gap in the present system, taking into account explicitly the angular dependence of the tunneling probability merely corresponds to a numerical prefactor. For anisotropic superconductors with $\vk$-dependent gaps, such an approximation is clearly not valid. Similarly to Ref.~\onlinecite{borkje}, one should then make the ansatz
\begin{equation}\label{eq:tun}
|\mathcal{T}_{\vk\vp}|^2 = \mathcal{T}^2 \sin\vartheta_\text{R}\sin\vartheta_\text{L} \Theta[\text{sgn}(\sin\vartheta_\text{R})\cdot\text{sgn}(\sin\vartheta_\text{L})],
\end{equation}
where $\mathcal{T}$ is a real constant, and the angles entering in Eq. (\ref{eq:tun}) define the trajectories of the quasiparticles involved in tunneling; see Fig. \ref{fig:tun} below. 

\begin{figure}[h!]
\centering
\resizebox{0.48\textwidth}{!}{
\includegraphics{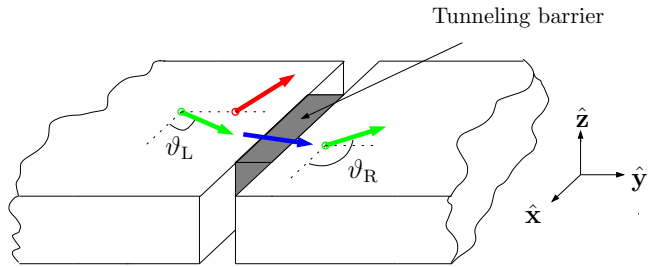}}
\caption{(Color) The tunneling scenario illustrated for two quasiparticles approaching the barrier separating the superconductors. For incoming momenta with a large component perpendicular to the barrier (green), tunneling occurs with greater probability than for incoming momenta with a small component perpendicular to the barrier (red). The sign of the component of momentum perpendicular to the barrier must be preserved in the process. For $s$-wave superconductors, the tunneling matrix element may be approximated by a constant, while it may not for anisotropic superconductors.}
\label{fig:tun}
\end{figure}

\begin{figure}[h!]
\centering
\resizebox{0.47\textwidth}{!}{
\includegraphics{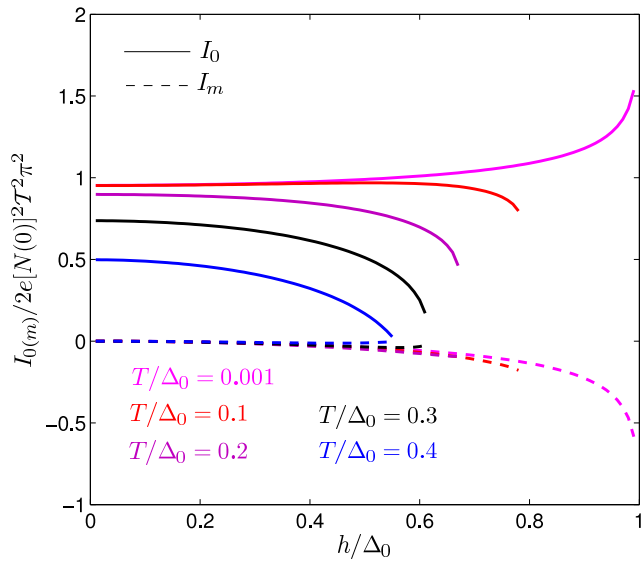}}
\caption{(Color) Plot of the components $I_0$ and $I_m$ as a function of exchange field $h$ for several temperature $T$. It is seen that $I_m$ becomes non-zero only as $h$ increases towards $\Delta_0$, such that the Josephson current is only sensitive to a rotation of the misorientation of the exchange fields in this regime.}
\label{fig:JosComp}
\end{figure}

\begin{figure}[h!]
\centering
\resizebox{0.47\textwidth}{!}{
\includegraphics{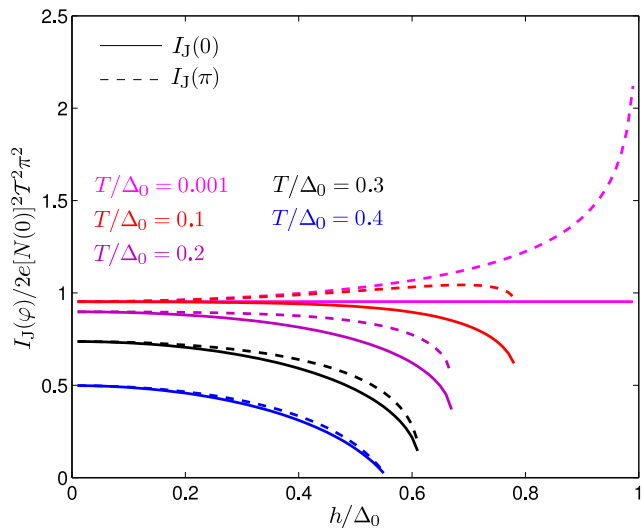}}
\caption{(Color) Plot of the total Josephson current in the parallell $I_\text{J}(0)$ and antiparallell $I_\text{J}(\pi)$ configuration of the exchange fields on both sides of the junction. It is seen that the Josephson current is actually enhanced with increasing field strength for the antiparallell configuration for low enough temperatures, in agreement with the result of Refs.~\onlinecite{li2002, bergeretJOS}.}
\label{fig:JosTot}
\end{figure}

\begin{figure}[h!]
\centering
\resizebox{0.47\textwidth}{!}{
\includegraphics{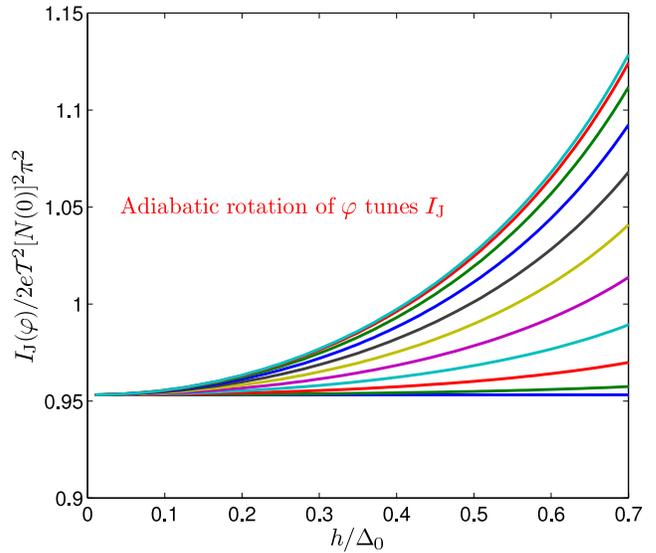}}
\caption{(Color) Plot of the total Josephson charge-current at $T/\Delta_0=0.001$ as a function of $h$ up to the critical field $h_c = 0.7\Delta_0$ in the presence of an adiabatic rotation of $\varphi$, ranging from $\varphi=0$ to $\varphi=\pi$ in steps of $0.1\pi$ from bottom to top. }
\label{fig:JosRotate}
\end{figure}

Having stated this, we are now able to investigate quantitatively how the Josephson charge-current in our system depends on the relative orientation of the exchange fields on both sides of the junction. The misalignment $\varphi$ of the exchange fields enters the expression for the Josephson charge-current through Eq. (\ref{eq:main}), which accounts for the qualitative behaviour. Converting the summation to integration as described above, we obtain
\begin{align}
I_0 &= 2e\mathcal{T}^2[N(0)]^2 \Delta(T,h)^2 \int^{\omega_0}_{-\omega_0}\int^{\omega_0}_{-\omega_0} F^+(\xi_1,\xi_2)\notag\\
&\times \prod_{i=1,2} \Bigg[\Big(1 + \frac{[\Delta(T,h)]^2}{\{\xi_i + \sqrt{\xi_i^2 + [\Delta(T,h)]^2}\}^2}\Big)^{-1}\notag\\
&\hspace{0.4in}\times  \frac{\text{d}\xi_i}{\xi_i + \sqrt{\xi_i^2 + [\Delta(T,h)^2]}}\Bigg] ,
\end{align}
while $I_m$ is given by the above expression by performing the substitution $F^+(\xi_1,\xi_2) \to F^-(\xi_1,\xi_2)$.
However, it is obvious that if $I_0 \gg I_m$, the effect of rotating $\varphi$ will be very small. For the purpose of obtaining a Josephson current which may be controlled by rotating the exchange fields, we are interested in obtaining $I_m$ as large as possible. To see if this is possible, we need to investigate under what circumstances varying
$\varphi$ will have an appreciable effect on the total Josephson current. Earlier works \cite{li2002, bergeretJOS} 
have considered a similar systems as the one considered in this paper, but restricted the exchange field orientations to be either parallel or antiparallel. Hence, our work represents a considerable extension of these results. Furthermore, we explicitly compute the relative magnitude between the term $I_m$, that provides the possibility 
of controlling $I_\text{J}$ by rotating $\varphi$, and the "intrinsic" Josephson-term $I_0$. Consider Fig. \ref{fig:JosComp} for a plot of $I_0/2e[N(0)]^2\mathcal{T}^2\pi^2$ and 
$I_m/2e[N(0)]^2\mathcal{T}^2\pi^2$, and  Fig. \ref{fig:JosTot}  for the total Josephson current $I_J$, as a 
function of $h/\Delta_0$ for several values of $T/\Delta_0$.
From Fig. \ref{fig:JosComp}, it is seen that $I_m$ is non-zero only when $h\to h_c$ for any temperature. This suggests that the Josephson current will only respond to a rotation of the exchange fields through the $I_m\cos\varphi$ term at very low temperatures and fields close to their critical values. Specifically, for the parallell and antiparallell configuration, this statement is consistent with the findings of Refs.~\onlinecite{li2002, bergeretJOS}. In general, however, we have here shown that an adiabatic rotation of $\varphi$ may offer a well-defined mechanism of tuning the magnitude of the Josephson current, as shown in Fig. \ref{fig:JosRotate}. One infers that the increase of $I_\mathrm{J}$ may be as large as 20\%. Note that the formal logarithmic divergence of the current in Fig. \ref{fig:JosTot} for $h\to \Delta_0$ when $T\to 0$ may be removed by considering higher orders of the tunneling matrix probability \cite{bergeretJOS}. Practically speaking, this divergence is clearly not of any concern since the critical field is determined by Tab. \ref{tab:list}, which states that $h_c/\Delta_0 \to 1/\sqrt{2}$ as $T\to0$.

\section{Discussion}\label{sec:dis}
\begin{figure}[h!]
\centering
\resizebox{0.48\textwidth}{!}{
\includegraphics{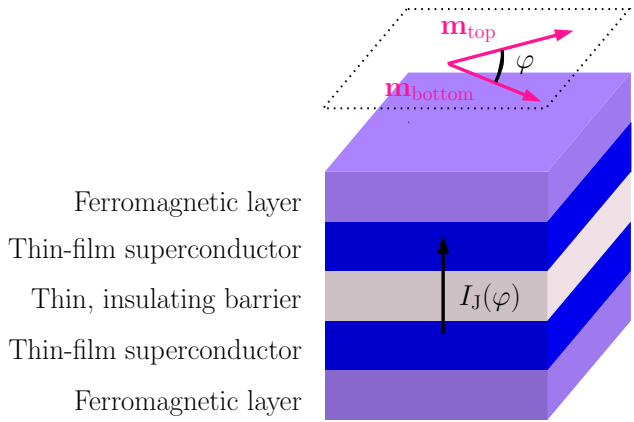}}
\caption{(Color) Suggested experimental setup for achieving homogeneous exchange fields in the superconductor. The antiparallell alignment of the exchange fields is probably the most viable to realize in order to avoid the Fraunhofer diffraction of the resulting Josephson current.}
\label{fig:SFlayer}
\end{figure}

A possible realization of the system proposed in the present 
paper could be achieved by either applying external magnetic fields to a thin-film S/I/S structure, or by considering two thin S/F bilayers with misaligned magnetization orientations separated by a thin, insulating barrier (see Fig. \ref{fig:SFlayer}). In such a geometry, the influence of the FM layers is non-local in the superconductor, such that the exchange field may be considered homogeneous \cite{bergeret2001}. Another important point concerns the thickness of the superconducting films, which would need to fulfill $d<\xi\ll\lambda$ in order for the exchange field to penetrate the film uniformly (note that the screening currents giving rise to the Meissner effect are suppressed in this geometry) \cite{meservey}, although making the film too thin could give rise to problems with $T_c$ being too small \cite{burnell}.
Moreover, it is likely that the Josephson current would display a Fraunhofer diffraction pattern if one cannot 
find a way of avoiding magnetic flux from the FM layers to penetrate the barrier. In this respect, the antiparallell alignment of the exchange fields is probably the most promising, since the flux penetration of the barrier could be expected to cancel out. Applying a field perpendicular to the stack would not give rise to a Fraunhofer diffraction pattern, but since the demagnization factor $n$ in such a geometry is close to 1, the critical field would be very small \cite{burnell}. Recall that the relation between an applied field $\mathbf{H}_\mathrm{a}$ and the field set up 
by the superconductor $\mathbf{H}_\mathrm{i}$ may be written as \cite{sudbo}
\begin{equation}
\mathbf{H}_\mathrm{i} = \frac{1}{1-n}\mathbf{H}_\mathrm{a}.
\end{equation}
In the present paper, we have studied the tunneling limit equivalent to a low transparency barrier. The effect of increasing the transparency of the barrier was treated within the Blonder-Tinkham-Klapwijk-formalism \cite{btk} in Ref.~\onlinecite{li2002}, where it was found that $I_\text{J}$ was no longer enhanced by increasing $h$, regardless 
of whether the orientation of the exchange fields was parallell or antiparallell. In the high transparency case, $I_\text{J}$ actually decreased more rapidly as a function of $h$ when $\varphi=\pi$ compared to $\varphi=0$. This shows that the Josephson-current would still be sensitive to a rotation of $\varphi$, although now the $\varphi=0$ configuration would correspond to the largest critical current.\\
\indent If an experimental setup as suggested here could be realized, the effect of the interplay between $\varphi$ and $\Delta\theta$ in $I_\text{J}$ may be observed in the following manner. For a superconductor-superconductor junction, the critical Josephson current is detected through the emission of microwave radiation with a power determined by the magnitude of the current and by the rate of change of the relative orientation between the exchange fields on both sides of the junction. This is the magnetic analogy of supplying an electrostatic potential to maintain an AC Josephson effect in the charge-channel. In this 
way, one maintains the novel AC oscillations both in the charge-Josephson current by rotating the exchange fields, even 
in the absence of an electrostatic voltage. Hence, a feasible experimental verification of the novel effect we predict in 
this paper would be the detection of microwave radiation 
associated with an AC Josephson effect originating with rotating magnetic field such that the misalignment angle varies with time. Note that rotating the fields on both sides of the junction with equal frequencies gives no AC effect.
\par
We close by reemphasizing that the above ideas should be experimentally realizable by \eg utilizing 
various geometries in order to vary the demagnetization 
fields.  Alternatively, one may use exchange biasing to 
an anti-ferromagnet by depositing an anti-ferromagnetic layer on top of the whole structure shown in Fig. \ref{fig:SFlayer}. Techniques of achieving
non-collinearity are routinely used in ferromagnet-normal metal structures \cite{bass1999}.

\section{Summary}\label{sec:sum}
In this paper, we have studied the Josephson charge-current that arises over a junction separating two thin-film $s$-wave singlet superconductors in the presence of misaligned in-plane exchange fields. A possible realization of such a system is visualized in Fig. \ref{fig:SFlayer}, where the idea is that a thin S/F layer may be considered as a superconductor with a homogeneous exchange field present \cite{meservey}.
The analytical solution within the Matsubara formalism reveals 
an interplay between the misorientation of the exchange fields, described by the angle $\varphi$, and 
the SC phase difference $\Delta\theta$ through the relation $I_\text{J} = (I_0 + I_m\cos\varphi)\sin\Delta\theta$, where $I_0$ and $I_m$ are real constants. Using a self-consistently obtained solution of the SC order parameter, we obtain a numerical plot of the Josephson current for arbitrary exchange fields and temperatures. Specifically, we examine the magnitude of $I_0$ and $I_m$ in order to investigate whether the $I_m$ term may contribute significantly to $I_\text{J}$ or not. While previous works have considered only the parallell ($\varphi=0$) or antiparallell ($\varphi=\pi$) configuration of the fields, our results show that the Josephson current will respond to any rotation of the orientation of the fields through the term $I_m\cos\varphi$. Consequently, we have analytically and numerically made an important distinction between the contributions to $I_\text{J}$ that stem from an "intrinsic" Josephson-term $I_0$ and the term $I_m$ that allows for a manipulation of the Josephson-current through a tuning of $\varphi$. This clarifies exactly how $I_\text{J}$ depends on the field orientations in any configuration. We find that $I_0$ and $I_m$ become comparable only for values of the exchange field close to the critical value. In this case, the Josephson charge-current may be enhanced by the presence of the exchange fields and controlled in a well-defined manner by adiabatically rotating the field directions on each side of the junction.

\section*{Acknowledgments}
J. L. gratefully acknowledges G. Burnell for very helpful comments with regard to experimental 
considerations, and E. K. Dahl for clarifying an important point concerning the superconducting-normal 
phase transition. This work was supported by the Norwegian Research Council Grants No. 157798/432 
and No. 158547/431 (NANOMAT), and Grant No. 167498/V30 (STORFORSK). The authors acknowledge Center 
for Advanced Study at The Norwegian Academy of Science and Letters for their hospitality during
the academic year 2006/2007.

\end{document}